%% file: main.tex
\documentclass[12pt]{iopart}
\expandafter\let\csname equation*\endcsname\relax
\expandafter\let\csname endequation*\endcsname\relax
\usepackage{physics}
\usepackage{orcidlink}
\usepackage{graphicx}
\usepackage{amsfonts}
\usepackage{cite}
\usepackage{float}

\newtheorem{theorem}{Theorem}
\newcommand{\nkraus}[0]{{m}} 
\newcommand{\nstates}[0]{{m}}
\newcommand{\krausop}[0]{{M}}
\newcommand{\group}[0]{{l}}
\newcommand{\ntargets}[0]{{n_t}}

\begin{document}

\title[Divide-and-Conquer Simulation of Open Quantum Systems]{Divide-and-Conquer Simulation of Open Quantum Systems}
\author{Thiago Melo D. Azevedo$^{1,*}$, Caio Almeida$^2$, Pedro Linck$^2$, Adenilton J. da Silva$^{1,*}$, Nadja K. Bernardes$^{2,*}$}

\address{$^1$ Centro de Informática, Universidade Federal de Pernambuco, Recife-PE, Brazil.}
\address{$^2$ Departamento de Física, Centro de Ciências Exatas e da Natureza, Universidade Federal de Pernambuco, Recife-PE, 50670-901, Brazil.}
\address{$*$ Corresponding authors.}
\ead{tmda@cin.ufpe.br, caio.carneiro@ufpe.br, 
pedro.linck@ufpe.br,
ajsilva@cin.ufpe.br and nadja.bernardes@ufpe.br}

\begin{abstract}

\noindent One of the promises of quantum computing is to simulate physical systems efficiently. However, the simulation of open quantum systems - where interactions with the environment play a crucial role - remains challenging for quantum computing, as it is impossible to implement deterministically non-unitary operators on a quantum computer without auxiliary qubits. The Stinespring dilation can simulate an open dynamic but requires a high circuit depth, which is impractical for NISQ devices. An alternative approach is parallel probabilistic block-encoding methods, such as the Sz.-Nagy and Singular Value Decomposition dilations. These methods result in shallower circuits but are hybrid methods, and we do not simulate the quantum dynamic on the quantum computer. In this work, we describe a divide-and-conquer strategy for preparing mixed states to combine the output of each Kraus operator dilation and obtain the complete dynamic on quantum hardware with a lower circuit depth. The work also introduces a balanced strategy that groups the original Kraus operators into an expanded operator, leading to a trade-off between circuit depth, CNOT count, and number of qubits. We perform a computational analysis to demonstrate the advantages of the new method and present a proof-of-concept simulation of the Fenna-Matthews-Olson dynamic on current quantum hardware.
\end{abstract}

\input{introduction2}

\input{dilation_techniques}

\input{state_preparation}

\input{combining_dilations}

\input{expanded_dilations}

\input{experiments}

\input{fmosimnisq}

\section{Conclusion}
\label{sec:conclusion}
The simulation of arbitrary open quantum systems with the Stinespring dilation requires circuits with a CNOT count of $O(\nkraus d^2)$ when we implement it as an isometry~\cite{iten2016quantum}. Algorithms with a probabilistic implementation of each Kraus operator, such as the Sz.-Nagy~\cite{Hu2020opendynamics} and the SVD dilations~\cite{schlimgen_svd_2022}, require $O(d^2)$ CNOTs to simulate an open quantum system. However, one does not obtain the complete dynamic as an output; we must combine the measurement statistics in a classical post-processing phase.

In this work, we proposed a mixed-state initialization algorithm that produces circuits with depth $O(\log \nkraus)$ when using $(n+1)(\nkraus+1)$ ancilla qubits or $O(\log \nkraus \log d)$ when restricted to $\nkraus+1$ ancilla qubits. We can combine the initialization strategy with an algorithm based on the Sz.-Nagy or SVD dilations to simulate an open quantum system, with success probability $1/\nkraus$ and circuit depth $O(d^2 + log(\nkraus))$. The mixed-state preparation strategy could also be used in the subroutines of quantum data compression~\cite{araujo2024schmidtquantumcompressor} and error correction strategies~\cite{obrienDensitymatrixSimulationSmall2017, kukulskiProbabilisticQuantumError2023, liQuantumErrorCorrection2023}.

We performed a computational experiment to show the proposed method's reduction in circuit depth compared to the Stinespring dilation. We also conducted a proof-of-concept experiment in a real quantum computer for the FMO dynamics to show that we can utilize the proposed method in current NISQ devices by following an optimization routine.

Additionally, we proposed an intermediate algorithm that provides a balanced approach between the Stinespring and the parallel dilation methods, with depth $O(\group d^2)$, that groups $\group$ Kraus operators into an expanded Kraus operator that we then dilate using any of the Sz.-Nagy / SVD algorithms. By grouping $\group$ Kraus operators, the number of actual dilations reduces to $\nkraus/\group$, which, in turn, decreases the number of ancilla qubits required and augments the success probability for implementing the simulation to $\group/\nkraus$. The grouping parameter also shows a trade-off between the CNOT count and circuit depth, in which the depth increases with $\group$ while the CNOT count decreases with $\group$.

Future work could explore approximation techniques for simulating open quantum systems on noisy devices. This strategy includes approximating unitary or isometry decompositions and selectively dilating only the most relevant Kraus operators. Additionally, further research could investigate alternative block-encoding methods for implementing Kraus operators or simulating dynamics that do not require dilation.

\section*{Acknowledgments}
This work is supported by research grants from CNPq, CAPES and FACEPE (Brazilian research agencies). 

\appendix
\input{mt_cswap}
\section*{References}

\bibliographystyle{unsrt}
\bibliography{refs}

\end{document}

%% file: introduction2.tex
\section{Introduction}

One of the promises of quantum computing is the simulation of quantum systems more efficiently than classical computers, as the quantum hardware would naturally use properties such as superposition and entanglement~\cite{feynman2018simulating}. Since the establishment of quantum computing, there have been many algorithm proposals for simulating quantum systems~\cite{nielsen2010quantum, lloyd1996simulators,kais2014quantum, kassal2011simulating, georgescu2014quantum, abrams1997simulation, 
king2018observation, bravyi2008quantum, wu2002polynomial, babbush2014adiabatic, 
omalley2016scalable, young2012finite, xia2018quantum, xia2018electronic, karra2016prospects}, but several deal with isolated quantum systems\cite{macedaDigitalQuantumSimulation2025,guoMitigatingErrorsAnalog2025,fausewehQuantumManybodySimulations2024,altmanQuantumSimulatorsArchitectures2021c,chandaRecentProgressQuantum2024}. However, no physical system is truly isolated from its environment, and a theory describing the dynamics of the open quantum system (OQS) is essential. The theory of OQS is responsible for explaining out-of-equilibrium properties of quantum systems~\cite{dattaQuantumTransportAtom2005}, providing a framework to describe measurements, and understanding and dealing with the detrimental effects of the unavoidable interaction between the system and its environment~\cite{rivasOpenQuantumSystems2012a}. Moreover, OQS have potential applications in reservoir engineering for quantum computation \cite{Verstraete} and as a resource for quantum information processing, such as quantum key distribution \cite{PhysRevA.83.042321}, quantum metrology \cite{PhysRevA.84.012103,PhysRevLett.109.233601}, quantum teleportation \cite{Laine14}, and quantum communication \cite{Bylicka14,RojasRojas2024nonmarkovianityin}. Nevertheless, the simulation of open quantum systems~\cite{delgado-granados_quantum_2024} is challenging for quantum computers because one cannot implement the non-unitary description as a unitary quantum operation in quantum hardware.

One strategy for simulating the evolution of open quantum systems is to consider it a part of the evolution of an expanded closed system, which is the case of the Stinespring Theorem~\cite{Stinespring1955, Ticozzi_2017}. Still, it has a high computational cost and is unsuitable for NISQ devices~\cite{preskill2018quantum}, as they can only run short-depth quantum circuits. Different block encoding decompositions dilate a single Kraus operator into a unitary matrix and are implemented probabilistically~\cite{Hu2020opendynamics, schlimgen_quantum_2021, schlimgen_svd_2022, Hu_2022_fmo,wang_simulating_2023,langerSzNagyFoiasHarmonic1972, schlimgen_quantum_2022}. These decompositions allow for the parallelization of the simulation, dilating one Kraus operator in each circuit, which reduces the circuit depth because one can execute the circuits in parallel. 

Parallelization of simulation enables current block encoding strategies that obtain shallower circuits that are better suited for NISQ devices than Stinespring dilation, as they only dilate one Kraus operator in each subcircuit~\cite{preskill2018quantum}. However, parallel block-encoding techniques do not produce the complete resulting quantum state on the quantum hardware. For the current block-encoding strategies, we can only obtain the statistics of the quantum map by classically combining the measurement outcomes from each circuit, but we do not realize the output state in the quantum hardware.

The main goal of this work is to produce a quantum circuit that outputs the quantum state of the evolution of an OQS in the hardware with the asymptotic depth of the block-encoding decompositions. To achieve this result, we utilize a divide-and-conquer strategy to prepare a mixed quantum state, a combination of $\nkraus$ initial quantum states, with depth $O(\log{\nkraus})$. Using this mixed-state preparation strategy, we combine the results of the block encoding methods to obtain the complete dynamic realized in the quantum hardware. 

We utilize Sz.-Nagy~\cite{Hu2020opendynamics} and SVD dilations~\cite{schlimgen_svd_2022} as block-encoding methods. We make this choice as they are exact decompositions. Furthermore, we introduce a grouping scheme for Kraus operators into expanded Kraus operators, where the limit cases (no grouping at all and grouping all of them in the same Kraus operator) can be seen as analogous to Stinespring and Sz.-Nagy, showing that we can obtain the two ideas from the same general algorithm. 
Increasing the number $\group$ of grouped Kraus operators reduces the CNOT count and the number of qubits necessary and increases the algorithm's success probability.  But increasing $\group$ also increases the circuit depth. Lastly, we do a computational analysis for the depth, CNOT count, and number of qubits for the methods discussed and a proof-of-concept experiment simulating Fenna-Matthews-Olson (FMO) dynamics~\cite{Hu_2022_fmo} in current NISQ devices.

We divide the rest of this work into the following sections. Section~\ref{sec:related_works} reviews the existing algorithms used in this work and updates their CNOT counts; Section~\ref{sec:mixed_state_prep} describes the proposed mixed state preparation strategy; Section~\ref{sec:combining_dilations} introduces the strategy for combining multiple non-unitary dilations for open quantum system simulation; Section~\ref{sec:grouping_dilations} proposes a balanced algorithm that combines two Kraus operator into an expanded Kraus operator, which we use in the block-encoding dilations; Section~\ref{sec:computational_analysis} contains a computational comparison of the depth, CNOT counts and qubits used between the algorithms proposed and the Stinespring dilation; Section~\ref{sec:FMO_real} describes a proof-of-concept experiment for the simulation FMO operators~\cite{Hu_2022_fmo} in current NISQ devices; Section~\ref{sec:conclusion} concludes the article and discusses perspectives for future works.

%% file: dilation_techniques.tex
\section{Dilation Techniques}
\label{sec:related_works}

In this section, we describe techniques for simulating the dynamics of an open quantum system by dilating the Kraus operators into unitaries. 
We focus on exact block-encoding methods for the simulation of Open Quantum Systems, mainly the Stinespring, Sz.-Nagy, and SVD dilations. However, there are in the literature approximate dilation techniques\cite{suriTwoUnitaryDecompositionAlgorithm2023,schlimgen_quantum_2021,schlimgen_quantum_2022 }. 

\subsection{Kraus Operators}
 A quantum channel $\Lambda$ is defined as a completely positive trace-preserving (CPTP) map that acts on the convex space of the density operators $\rho$ of a given Hilbert space $\mathcal{H}$ with dimension $d$. Any quantum dynamic can be represented by a set of $\nkraus \leq d^2$ linear operators $\{\krausop_k\}_{k=1}^{\nkraus}$, where each element of the set is known as a Kraus operator, in such a way that the following equations hold \cite{rivasOpenQuantumSystems2012a, Hu2020opendynamics, petruccione2007opensystems}:
\begin{equation}\label{eq:krauseq}
\Lambda(\rho) = \sum_k \krausop_k\rho\krausop_k^\dagger,
\end{equation}
\begin{equation}\label{eq:krauscond}
\sum_{k}\krausop_k^\dagger \krausop_k = I.
\end{equation}

\par Note that $M_K$ is not necessarily a unitary operator. However, it can be seen as the part of a unitary evolution $U$ in a bigger closed system $\mathcal{H}\otimes \mathcal{H}_{env}$ \cite{petruccione2007opensystems,rivasOpenQuantumSystems2012a}, obtained after doing a closed evolution and tracing out the degrees of freedom associated to the environment Hilbert's space $\mathcal{H}_{env}$. The space $\mathcal{H}_{env}$ and the unitary $U$ are not uniquely determined, so there are different unitary representations of the same quantum channel.

\subsection{Stinespring Dilation}

\par The Stinespring Representation Theorem constructs a unitary $U_{St}$ for a quantum dynamic $\Lambda$ such that

\begin{equation}\label{eq:stinespringreptheorem}
        \Lambda(\rho) = \Tr_{E} (U_{St} (\rho \otimes \ket{0}\bra{0})  U^{\dagger}_{St}).
\end{equation}

\noindent For the sake of simplicity, the environmental state was fixed to $\ket{0}\bra{0}$. To obtain a unitary $U_{St}$ for a given set of Kraus operators $\{M_k\}_{k = 1}^\nkraus$, we impose
\begin{equation}\label{eq:stinespring_condition}
U_{St}(\ket{\psi} \otimes \ket{0} ) = \sum_{k=1}^{\nkraus}  \krausop_k \ket{\psi} \otimes \ket{k} .
\end{equation}
This determines the first $d$ columns of $U_{St}$, and we complete the other columns of the unitary using the Gram-Schmidt process
\begin{equation}\label{eq:ustinespring}
    U_{St} = \begin{bmatrix}
        \krausop_1 & \cdots & \cdots \\
        \krausop_2 & \cdots & \cdots \\
        \vdots & \dots & \vdots \\
        \krausop_{\nkraus} & \ddots & \dots \\
    \end{bmatrix}.
\end{equation}

\begin{figure}[H]
    \centering
    \includegraphics[scale=1.2]{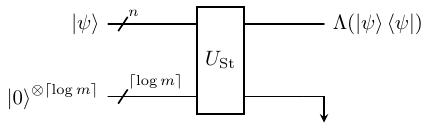}
    \caption{Implementation of a Stinespring dilation as a quantum circuit.  The implementation utilizes $ \lceil \log m \rceil$ auxiliary qubits for the dilation, that are traced out at the end to obtain the state $\Lambda (\ket{\psi} \bra{\psi})$ for the $n$ qubit system. }
    \label{fig:stinespring-circuit}
\end{figure}

\noindent The required condition from Eq.~\ref{eq:stinespring_condition} implies that $U_{St}$ fulfills the desired open dynamic after tracing out the auxiliary (environment) qubits. 
The circuit representation of the Stinespring unitary is shown in Fig.~\ref{fig:stinespring-circuit}, and utilizes $\lceil \log \nkraus \rceil$ auxiliary qubits to implement the dilated unitary~\cite{suriTwoUnitaryDecompositionAlgorithm2023}.

As the auxiliary register is initialized in state $\ket{0}$, we can implement the Stinespring decomposition as an isometry. In this way, we implement only the first $d$ columns of $U_{St}$ (columns containing the Kraus operators), reducing the depth and number of gates in the circuit implementation compared to the whole unitary. From Ref.~\cite{iten2016quantum}, the column-by-column decomposition for an isometry from $\beta$ qubits to $\alpha\geq \beta$ qubits has an upper bound of the CNOT count of $2^{\beta+\alpha}-\frac{1}{24}2^\alpha+O(\alpha^2)2^\beta$, where $2^\beta$ corresponds to the number of columns of the isometry and $2^\alpha$ to the number of lines. The upper bound is true for $\alpha\geq8$. 

For Stinespring decomposition, the $n$-qubit Hilbert space $\mathcal{H}$ has a dimension of $d = 2^n$, thus the isometry dimension is $md \times d$. Hence, the number of CNOT gates of the decomposition is $md^{2}-\frac{1}{24}md+O(\log^2_2 (md))d$ = $O(md^{2})$. The use of column-by-column isometry allows for a reduction in the gate cost of Stinespring dilation compared to previous implementations~\cite{suriTwoUnitaryDecompositionAlgorithm2023, Hu2020opendynamics}. The Stinespring dilation is deterministic and requires only one circuit call to be successfully implemented~\cite{suriTwoUnitaryDecompositionAlgorithm2023}.

\subsection{Parallel Sz.-Nagy Dilations }
\label{sec:parallelsznagy}
The work of~\cite{Hu2020opendynamics} introduced a quantum algorithm for simulating open quantum systems on quantum devices based on Sz.-Nagy dilation theorem~\cite{langerSzNagyFoiasHarmonic1972}. Given a quantum dynamic $\Lambda$ represented by a set of Kraus operators $\{\krausop_k\}_{k=1}^\nkraus$, the method utilizes the Sz.-Nagy Theorem to dilate each Kraus operator into a unitary matrix, so that each Kraus operator can be implemented individually in parallel. This allowed the authors to significantly reduce the circuit depth compared to previous methods based on the Stinespring dilation. However, a significant limitation is that the method does not output the complete simulation, only a single Kraus operator $\krausop_k$ individually, with the measurement statistics being combined in a classical post-processing phase. One auxiliary qubit is needed to construct the unitary, and the Kraus operator is only implemented after a selective measurement; hence, it is a probabilistic implementation.

Suppose a given quantum dynamic is represented by a set of Kraus operators $\{\krausop_k\}$ acting on the Hilbert space $\mathcal{H}$. For each Kraus operator $\krausop_k$, there exists a unitary $U^{SN}_k$ that acts on an expanded Hilbert space containing a single additional qubit defined by

\begin{equation}\label{eq:sznagy_def}
    U^{SN}_{k} = \begin{bmatrix}
        \krausop_k & D_{\krausop_{k}} \\
        D_{\krausop_{k}^{\dagger}} & -\krausop_k^{\dagger} \\
    \end{bmatrix},
\end{equation}
where each operator $D_T = \sqrt{I - T^\dagger T}$ is called the defect operator for the operator $T$. 
\par The action of this unitary in the quantum state $\ket{\psi}\otimes \ket{0}$ can be interpreted as

\begin{equation}\label{eq:action_sznagy}
    U_{k}^{SN} (\ket{\psi}\otimes \ket{0}) = \krausop_k\ket{\psi} \otimes \ket{0} +  D_{\krausop_{k}^{\dagger}} \ket{\psi} \otimes \ket{1}.
\end{equation}

\begin{figure}[H]
    \centering
    \includegraphics[scale=1.2]{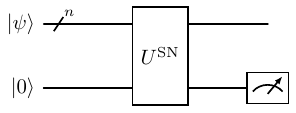}
    \caption{Implementation of a Sz.-Nagy dilation as a quantum circuit. The auxiliary qubit is measured, and if the measurement outcome is $\ket{0}$, we obtain $\krausop_k\ket{\psi}$ in the $n$ system qubits. }
    \label{fig:sznagy-circuit}
\end{figure}

\noindent The operator $\krausop_k$ is successfully implemented if the ancillary register is measured in the state $\ket{0}$, with probability $p_k = || \krausop_k \ket{\psi} ||^2$. This requires $O(1/p_i)$ circuit runs to be implemented successfully. The total number of circuit calls expected to implement all Kraus operators is equal to or greater than $\nkraus^2$~\cite{suriTwoUnitaryDecompositionAlgorithm2023}. A circuit implementation of the Sz.-Nagy dilation can be seen in Fig.~\ref{fig:sznagy-circuit}.

However, this method can be parallelized and executed on multiple systems simultaneously, which reduces the simulation running time. In this case, the running time is bounded in the upper order by the largest value of $1/p_i$, i.e., the operator with the lowest probability.

For each Kraus operator $\krausop_k$, we can implement the Sz.-Nagy dilation as a column-by-column isometry~\cite{iten2016quantum}, implementing only the first $d$ columns containing $2$ operators that act on a $n$-qubit space with dimension $d=2^n$. The isometry will have $2d$ lines and $d$ columns, thus, the number of CNOT gates is given by $2d^2-\frac{1}{24}d+O(\log^2_2 (2d) )d$, in which $2d^2$ is the leading term. If the $\nkraus$ operators were implemented in the same circuit, the total CNOT count would have a leading term of $2\nkraus d^2$. However, since they are implemented in parallel, each dilation would have a CNOT count by $2d^2$, leading to a reduced depth compared to the Stinespring dilation.

\subsection{Singular Value Decomposition Dilation}
\label{sec:svd_dilation}

In Ref.~\cite{schlimgen_svd_2022}, the authors introduced an implementation for non-unitary operators using the classical Singular Value Decomposition (SVD) and one ancilla qubit. Given a non-unitary operator $\krausop_k$, the operator is decomposed using SVD into a non-unitary diagonal matrix $\Sigma_k$ and unitary matrices $U_k, V_k^\dagger$ as follows

\begin{equation}\label{eq:svd_def}
    \krausop_k = U_k \Sigma_k V_k^\dagger.
\end{equation}

\noindent We assume that the non-unitary diagonal operator $\Sigma_{k}$ has complex diagonal entries $\sigma_{kjj}$, and the magnitude of each element is equal or less than one~\cite{schlimgen_svd_2022}. The non-unitary diagonal can be implemented using a dilated unitary diagonal defined as

\begin{equation}\label{eq:usigma_def}
    U_{\Sigma_k} = \begin{pmatrix}
        \Sigma_{k}^{(+)} & 0 \\
        0 & \Sigma_{k}^{(-)}
    \end{pmatrix},
\end{equation}

\noindent where $\Sigma_k = \frac{1}{2}(\Sigma_{k}^{(+)} +  \Sigma_{k}^{(-)})$. Each element of the diagonal dilation can be obtained by~\cite{schlimgen_svd_2022}
\begin{equation}\label{eq:sigmapm_def}
    \Sigma_{kjj}^{(\pm)} = \sigma_{kjj} \pm i\sqrt{1-||\sigma_{kjj}||^2}.
\end{equation}

\begin{figure}[H]
    \centering
    \includegraphics[scale=1.2]{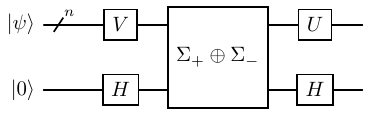}
    \caption{Circuit implementation of the Singular Value Decomposition dilation for a non-unitary operator $\krausop_k$. The non-unitary operator is decomposed into $\krausop_k = U_k \Sigma_k V^{\dagger}_k$, where $U_k, V_k$ are unitaries and $\Sigma_k$ is a non-unitary diagonal. $\Sigma_k$ is then dilated into a unitary $U_{\Sigma_k} = \Sigma_{k}^{(+)} \oplus \Sigma_{k}^{(-)}$, which acts on the system qubits and in an ancilla qubit.}
    \label{fig:svd-circuit}
\end{figure}

This decomposition can be used for the Kraus operators $\{\krausop_k\}_{k=1}^\nkraus$ representing a given quantum dynamic $\Lambda$. Each Kraus operator is decomposed into the SVD form with a singular value matrix $\Sigma_{k}$ and then dilated to $U_{\Sigma_k}$ using an ancilla qubit~\cite{schlimgen_svd_2022}. The circuit implementation can be seen in Fig.~\ref{fig:svd-circuit}.

The action of the dilation algorithm in the state initialized in $\ket{\psi} \otimes \ket{0}$, where $\ket{\psi}$ is a $n$-qubit state, is given by
\begin{equation}\label{eq:svd_action}
    \begin{aligned}
        \ket{\Psi} &= (U_k\otimes H) (\Sigma_{k}^{(+)} \oplus \Sigma_{k}^{(-)}) (V_k^\dagger \otimes H) (\ket{\psi}\otimes \ket{0})\\
                   &= \krausop_k\ket{\psi} \otimes \ket{0} + \Tilde{\krausop}_k \ket{\psi} \otimes \ket{1},
    \end{aligned}
\end{equation}

\noindent where $\tilde{\krausop}_k = U_k(\Sigma_{k}^{(+)} - \Sigma_{k}^{(-)})V_{k}^\dagger $. 

The implementation is successful when the ancilla is measured in the state $\ket{0}$, which happens with probability $p_k = ||\krausop_k \ket{\psi}||^2$, similarly to the Sz.-Nagy dilation. It also only dilates a single Kraus operator, not realizing the simulation of the entire quantum map in the quantum hardware.

An $n$-qubit unitary gate can be decomposed into up to $(23/48) \times 4^n - (3/2) \times 2^n + 4/3$ CNOT gates, using the version of the Quantum Shannon
Decomposition~\cite{shende2006circuits} with the lowest gate count. Considering that we have two $n$-qubit unitary gates and that $2^n=d$, then they would contribute with $(23/24)  d^2 - 3 d + 8/3$ CNOTs. This CNOT cost is lower than the $O(n^2 2^{2n})$ gate count reported in Ref.~\cite{schlimgen_svd_2022}. Meanwhile, the unitary diagonal gate $U_\Sigma$ acting on the $n+1$ qubits can be decomposed into up to $2^{n+2}-3 = 4 d -3$ elementary gates~\cite{bullock2004diagonal}. Hence, the asymptotic gate cost of this decomposition is $O(d^2)$, similar to the Sz.-Nagy decomposition when implemented using an isometry, but because of the lower constants, the SVD decomposition achieved lower depths and gate counts in our computational implementation compared to Sz.-Nagy dilations with isometry. Therefore, in the following sections, we choose the SVD algorithm for the dilation of Kraus operators.

%% file: state_preparation.tex
\section{Mixed State Preparation}
\label{sec:mixed_state_prep}

In this section, we utilize a strategy to prepare a mixed quantum state by combining multiple initialized states with CSWAPs controlled by ancilla qubits. 
It is a divide-and-conquer strategy~\cite{araujoDivideandconquerAlgorithmQuantum2021}, as we initialize individual components and then build the final solution from them.

\subsection{Divide-and-Conquer algorithm}
Given two $n$-qubit systems prepared in the states $\rho_{1}$ and  $\rho_{2}$, respectively,  and an ancilla qubit $\ket{p} = \sqrt{p_1} \ket{0} + \sqrt{p_2} \ket{1}$. We assume that initially, the total system $\rho$ is prepared as

\begin{equation}\label{eq:initial_state}
    \rho= \rho_{1}\otimes \rho_{2} \otimes \ket{p} \bra{p}.
\end{equation}

For two states $\rho_{1}, \rho_{2}$, our aim is to prepare the state $\rho_m = p_1\rho_{1} + p_2\rho_{2}$, using a CSWAP with the ancilla qubit $\ket{p} = \sqrt{p_1} \ket{0} + \sqrt{p_2} \ket{1}$ acting as a control. The application of multi-qubit CSWAPs in the first two systems, which are controlled by the ancilla system, results in

\begin{equation}
\begin{aligned}
\rho' &= \text{CSWAP} (\rho_{1}\otimes \rho_{2}\otimes \ket{p} \bra{p} )\text{CSWAP}^{\dagger} \\
&= p_1\rho_{1} \otimes \rho_{2}\otimes \ket{0} \bra{0} + \sqrt{p_1 p_2}\rho_{1}\otimes\rho_{2} \text{SWAP}^{\dagger}\ket{0}\bra{1} \\
&\quad + \sqrt{p_1 p_2} \text{SWAP}\rho_{1}\otimes\rho_{2}\otimes\ket{1} \bra{0} + p_2\rho_{2} \otimes\rho_{1} \otimes \ket{1} \bra{1},
\end{aligned}
\label{eq:cswap_rho}
\end{equation}
\noindent where $\text{SWAP}(\ket{\psi}\otimes\ket{\phi}) = \ket{\phi}\otimes\ket{\psi}$. After tracing out the two last subsystems, we obtain the desired state as follows

\begin{equation}
\begin{aligned}
    \text{Tr}_{23}(\rho') &= \text{Tr}_{2} \left(\bra{0}_{3}\rho'\ket{0} _{3}+\bra{1}_{3}\rho'\ket{1} _{3} \right) \\
&= \text{Tr}_2 \left(p_1\rho_{1} \otimes \rho_{2} +p_2\rho_{2}\otimes\rho_{1} \right)\\
&= p_1\rho_{1}\text{Tr}_{2}(\rho_{2})+p_2\rho_{2}\text{Tr}_{2}(\rho_{1}) \\
&= p_1\rho_{1} + p_2\rho_{2}.
\end{aligned}
\label{eq:mixed_twostate}
\end{equation}

\begin{figure}[H]
    \centering
    \includegraphics[scale=1.2]{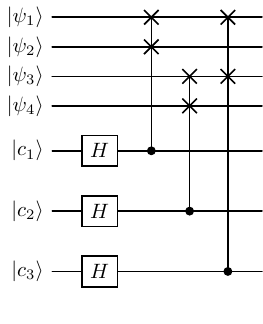}
    \caption{Example of a circuit that prepares the mixed state $\rho_f = \sum^4_{i=1} \ket{\psi_i}\bra{\psi_i}$, composed of $4$ pure states.}
    \label{fig:mixed-state}
\end{figure}

We can prepare a mixed state that is a combination of $\nstates=2^k$ using this strategy repeatedly, pair by pair. For example, given two mixed states $\gamma_{1} =  \frac{1}{2}\rho_{1}+ \frac{1}{2}\rho_{2}$ and $\gamma_{2}=\frac{1}{2}\rho_{3}+\frac{1}{2}\rho_{4}$ constructed with this strategy, we can use the strategy again for the states $\gamma_1$, $\gamma_2$, resulting in  

\begin{equation}
    \rho_m = \dfrac{1}{2}\gamma_{1} + \dfrac{1}{2}\gamma_{2} = \frac{1}{4}(\rho_{1}+\rho_{2}+\rho_{3}+\rho_{4}).
\end{equation}
A circuit implementation for preparing a mixed state from $4$ evenly distributed pure states is shown in Fig.~\ref{fig:mixed-state}.

Throughout the rest of this work, we use the described divide-and-conquer strategy to prepare mixed states for simulating open quantum systems, where we deal with evenly distributed states ($p_i = 1/\nstates, \forall i$). We can set $\ket{p} = H\ket{0} = \frac{1}{\sqrt{2}}\ket{0}+\frac{1}{\sqrt{2}}\ket{1}$. If we were dealing with an uneven distribution of the states, the ancilla qubit $\ket{p}$ would need to be adjusted for each CSWAP step.

\subsection{Computational cost}
To prepare a mixed state generated from $\nstates = 2^k$ states as described above, the method requires $k = \log_2 \nstates$ layers of CSWAPs acting on systems with $n$ qubits. Here, we consider two strategies for implementing the multi-qubits CSWAPs. First, we assume that we have one single auxiliary qubit to be used as a control for each multi-qubit CSWAP, so we can use the implementation described in \ref{sec:mtcswap}, which leads to a depth of $6 \lceil \log_2 n \rceil+14$ for each multi-qubit CSWAP. Each auxiliary qubit is initialized in the state $H\ket{0} = \frac{1}{\sqrt2}\ket{0}+\frac{1}{\sqrt2}\ket{1}$ and in each layer all the multi-qubit CSWAPs act in parallel. Since the multi-qubit CSWAP gates act in parallel, and there are $L =  \log_2 \nstates $ layers, the total depth of this circuit will be $  \log_2 \nstates (6 \lceil \log_2 n \rceil + 14)$, while the number of auxiliary qubits will be $\frac{\nstates}{2} + \frac{\nstates}{4} + \dots +\frac{\nstates}{\nstates} = \nstates-1$.

We could further reduce the depth by utilizing additional clean ancillas for each qubit targeted by the CSWAPs, instead of using a multi-qubit CSWAP. By preparing the ancilla in the entangled state $\ket{p} = \frac{1}{\sqrt{2}}\ket{0^{\otimes n}}+\frac{1}{\sqrt{2}}\ket{1^{\otimes n}}$, we can achieve the mixed state $\rho_m$. This will require $n(\nstates-1)$ auxiliary ancillae, but every CSWAP is implemented in parallel, leading to a circuit depth of  $14  \log_2 \nstates $.

%% file: combining_dilations.tex
\section{Combining the non-unitary dilations}
\label{sec:combining_dilations}

In this section, we aim to use the mixed-state preparation method, introduced in Section~\ref{sec:mixed_state_prep}, to combine the action of individually dilated Kraus operators. The algorithm used to dilate each Kraus operator may be chosen between the SVD and the Sz.-Nagy Dilation. This implementation simulates the whole quantum dynamic in the quantum hardware and does not require a classical post-processing phase. \\

For each Kraus operator $M_k$, the resulting state after applying the unitary on the initial state $\ket{\psi}\otimes\ket{0}$ is given by

\begin{equation}
     \ket{\Psi_{k}} =\krausop_{k}\ket{\psi}\otimes \ket{0} + \Tilde{D}_{\krausop_{k}}\ket{\psi}\otimes \ket{1}, 
\end{equation}
where $\tilde{D}_{\krausop_{k}}$ is the defect operator.

\par After applying the divide-and-conquer algorithm for all possible states $\{\ket{\Psi_k}\}_{k=1}^m$ and tracing out the remaining registers, the resulting state will be
\begin{equation}\label{eq:state_postmix}
    \begin{aligned}
        \rho_m & = \frac{1}{\nkraus}\sum_{k}^{m}  \ket{\Psi_{k}} \bra{\Psi_{k}}\\
        & =\frac{1}{\nkraus}\sum_{k}^{\nkraus} \big(  M_{k}\ket{\psi} \bra{\psi} M_{k}^{\dagger} \otimes \ket{0} \bra{0}+  M_{k}\ket{\psi} \bra{\psi} \tilde{D}_{\krausop_k}\otimes \ket{0} \bra{1} \\
           &+ \tilde{D}_{\krausop_k}\ket{\psi} \bra{\psi} M_{k}\otimes  \ket{1} \bra{0} + \tilde{D}_{\krausop_k}\ket{\psi} \bra{\psi} \tilde{D}_{\krausop_k}\otimes \ket{1} \bra{1} \big).
    \end{aligned}
\end{equation}

\par To obtain the result of the desired quantum channel, it is necessary to apply a measurement on the ancilla qubit, selecting only $\ket{0}$. The final state is given then by

\begin{equation}\label{eq:select_measure}
    \begin{aligned}
        \rho_f 
& = \sum_{k}^{\nkraus}M_{k}\ket{\psi} \bra{\psi} M_{k}^{\dagger}.
    \end{aligned}
\end{equation}

\par The simulation is only successfully implemented when we measure the state $\ket{0}$ in the auxiliary qubit. Therefore, the success probability $P_0 = P(\ket{0})$ is

\begin{equation}
    \begin{aligned}
        P_0 &= \Tr(\ket{0} \bra{0} \rho_m\ket{0} \bra{0} ) \\
&= \frac{1}{\nkraus}\Tr\left(\sum_{k}^{\nkraus}M_{k}\ket{\psi} \bra{\psi} M_{k}^{\dagger}\right)= \frac{1}{\nkraus},
    \end{aligned}
\end{equation}
\noindent independently of the chosen dilation method.

Since we combine $\nkraus$ states of dimension $d$, we have $\log_2 \nstates$ layers of CSWAPs. Implementing the proposed mixing method adds a cost equivalent to combining $m$ states of $n=\log_2 d$ qubits as proposed in Section~\ref{sec:mixed_state_prep}. Therefore, combining the $\nkraus$ dilated states adds a circuit depth of $\log_2 \nkraus (6 \lceil \log_2 \log_2 d \rceil + 14)$, if we use $\nkraus-1$ ancillae, or $14\log_2 \nkraus$, when using $(\nkraus-1)(n+1)$ ancillae. Since $\nkraus\leq d^2$ and the dilations contribute with a circuit depth $O(d^2)$, the asymptotic circuit depth of the complete circuit will be $O(d^2)$.

The successful implementation of the combined Sz.-Nagy or SVD dilations will require $O(\frac{1}{P_0}=m)$ runs of a circuit with a depth of $O(d^2)$. Meanwhile, the Stinespring dilation has a circuit depth of $O(md^2)$. This results in the same $O(md^2)$ total computational resources; however, SVD or Sz.-Nagy dilations result in a shallower circuit depth of the actual implementation. This amounts to the exchange of quantum execution time for classical execution time, while the former is associated with noise and decoherence effects in the quantum hardware.

%% file: expanded_dilations.tex
\section{Grouping Kraus Operators}
\label{sec:grouping_dilations}

Until now, we have focused on dilating each Kraus operator individually. However, it is also possible to dilate multiple Kraus operators simultaneously. This is particularly interesting because, in the maximal grouping, we obtain a result similar to the Stinespring technique described previously, and in the minimal grouping, the Sz.-Nagy algorithm, meaning that both dilations are manifestations of the same dilation scheme.

 To achieve this, we define an expanded Hilbert space and obtain new, expanded Kraus operators. Defining a Hilbert space $\mathcal{H}$ with dimension $\tilde{d} = \group d$, we can group up $\group$ of the original Kraus operators $\krausop_k$ in the first $\group d$ columns of the expanded Kraus operators $\krausop_k$. To guarantee that $\sum_k \krausop^{\dagger}_k \krausop_k = I$, we also define a final expanded Kraus operator, which contains only the identity matrix in the second column.

The construction of the operators can be generalized for the grouping of $\group$ Kraus operators into an expanded operator. Using long division, we can write the number of Kraus operators as $\nkraus = \group\cdot b +r $, $r,b \in \mathbb{Z}$ with $0\leq r < l$, which can be grouped as follows

\begin{equation}\label{eq:u-stinespring}
    \tilde{\krausop_1} = \begin{bmatrix}
        \krausop_1 & 0 &\cdots & 0 \\
        \krausop_2 & 0 & \cdots & 0 \\
        \vdots & 0 & \dots & 0 \\
        \krausop_{\group} & 0 & \ddots & 0 \\
    \end{bmatrix}, \tilde{\krausop_2} = \begin{bmatrix}
        \krausop_{\group+1} & 0 &\cdots & 0 \\
        \krausop_{\group+2} & 0 & \cdots & 0 \\
        \vdots & 0 & \dots & 0 \\
        \krausop_{2\group} & 0 & \ddots & 0 \\
    \end{bmatrix},\dots, 
    \tilde{\krausop_b} = \begin{bmatrix}
        \krausop_{(b-1)\group+1} & 0 &\cdots & 0 \\
        \krausop_{(b-1)\group+2} & 0 & \cdots & 0 \\
        \vdots & 0 & \dots & 0 \\
        \krausop_{bl} & 0 & \ddots & 0 \\
    \end{bmatrix}
\end{equation}
If $r > 0$, we form a group such as
\begin{equation}
\tilde{\krausop}_{b+1} = \begin{bmatrix}
        \krausop_{b\group+1} & 0 &\cdots & 0 \\
        \vdots & 0 & \cdots & 0 \\
        \krausop_{b\group+r} & 0 & \dots & 0 \\
        0 & 0 & \ddots & 0 \\
        \vdots & \vdots & \ddots & 0 \\
        0 & 0 & \ddots & 0 \\

    \end{bmatrix},
\end{equation}
To ensure that this new set of operators represents a completely-positive and trace-preserving (CPTP) map, it is also necessary to implement

\begin{equation}
    \tilde{\krausop}_{b+2} = \begin{bmatrix}
        0 &  0 & \cdots & 0  \\
        0 & I & \vdots & \vdots  \\
        \vdots & \cdots & \ddots & 0 \\
        0 & 0 & \cdots & I  \\
        \end{bmatrix}
\end{equation}
The top left of $\tilde{\krausop}_{b+2}$ is a block of zeros of size $d\times d$, while the rest of the lines and columns are equal to an identity of dimension $(l-1)d\times(l-1)d$.

Now we explore an example of this procedure. For a set of $4$ Kraus operators $\{\krausop_1, \krausop_2, \krausop_3, \krausop_4\}$, we group $2$ operators into a single expanded $\tilde{\krausop}_k$, which results in
\begin{align}\label{eq: exp_ops_list}
    \tilde{\krausop}_1 = \begin{bmatrix}
    \krausop_1 & 0 \\
    \krausop_2 & 0 
\end{bmatrix}, \quad
\tilde{\krausop}_2 = \begin{bmatrix}
    \krausop_3 & 0 \\
    \krausop_4 & 0 
\end{bmatrix}, \quad
\tilde{\krausop}_3 = \begin{bmatrix}
    0 & 0 \\
    0 & I 
\end{bmatrix}.
\end{align}

Where it is possible to see that they still represent a CPTP mapping due to $    \tilde{\krausop_1}^{\dagger}\tilde{\krausop_1} +  \tilde{\krausop_2}^{\dagger}\tilde{\krausop_2} +  \tilde{\krausop_3}^{\dagger}\tilde{\krausop_3} = I$. Applying the quantum channel,

\begin{equation}
\begin{aligned}
\Lambda(\rho)&=\Tilde{\krausop_{1}}(\rho)\Tilde{\krausop_{1}}^{\dagger} + \Tilde{\krausop_{2}}\rho \Tilde{\krausop_{    2}}^{\dagger} + \Tilde{\krausop_{3}}\rho \Tilde{\krausop_{3}}^{\dagger} \\
&=  \krausop_{1}\rho \krausop_{1}^{\dagger} \otimes \ket{0} \bra{0} +  \krausop_{2}\rho \krausop_{2}^{\dagger} \otimes \ket{1} \bra{1} + \krausop_{3}\rho \krausop_{3}^{\dagger} \otimes \ket{0} \bra{0} + \krausop_{4}\rho \krausop_{4}^{\dagger} \otimes \ket{1} \bra{1}. 
\end{aligned}
\end{equation}

 \noindent Note that it is still possible to obtain the original map by tracing out the auxiliary subspace

 \begin{equation}
         \Tr_{2}\Lambda(\rho) = \krausop_{1}\rho \krausop_{1}^{\dagger} + \krausop_{2}\rho \krausop_{2}^{\dagger} + \krausop_{3}\rho \krausop_{3}^{\dagger} + \krausop_{4}\rho \krausop_{4}^{\dagger}.
 \end{equation}
 
\par This technique may be of interest in cases where the depth and gate count of the mixed-state preparation with CSWAPs can be a significant part of the total depth and gate count of the circuit or if one needs to reduce the number of qubits used. Furthermore, reducing the number of CSWAPs to be implemented will also increase the probability of success of implementation.

To implement the set of expanded operators, we can use any of the dilation methods described in Sections~\ref{sec:parallelsznagy} and~\ref{sec:svd_dilation}, replacing $\krausop_k$ by $\Tilde{\krausop}_k$. For both strategies, if we group $\group$ Kraus operators, the number of non-unitary dilations is reduced from $\nkraus$ to $\nkraus/\group$. This results in an increase of the probability from $1/\nkraus$ to $\group/\nkraus$.

\subsection{Group Sizes and Computational Cost}

We now focus on the effects of varying group sizes on the CNOT count and circuit depth of, respectively, each of the dilation methods and the mixing step described in Section~\ref{sec:mixed_state_prep}. Furthermore, we assume from now on, without loss of generality, that both $m$ and $l$ are powers of 2.
\par The CNOT count of each dilation method can be obtained by replacing $d \rightarrow ld$ in Sections \ref{sec:parallelsznagy} and \ref{sec:svd_dilation}. This results in the isometry implementation of the Sz.-Nagy dilations having a lower asymptotic cost of $O(\group d^2)$ in comparison to the SVD dilations, which has $O(\group^2 d^2)$. However, it is important to note that, for $\group=2$, the SVD decomposition still yielded a lower circuit depth when compared to applying the Sz.-Nagy dilation as an isometry.

Due to the decrease in the number of actually implemented dilations, which went from $\nkraus$ to $\nkraus/\group$, the cost of the mixing step decreases, as can be explicitly obtained by replacing $m \rightarrow m/l $ in the results of Section~\ref{sec:mixed_state_prep}. The final asymptotic cost combines the dilation and mixing cost, resulting in a total circuit depth of $O(ld^2)+O(\log \nkraus-\log \group) $, assuming $(m/l - 1)(n+1+\log l)$ ancillae are available. 

\begin{figure}[H]
    \centering
    \includegraphics[width = \linewidth]{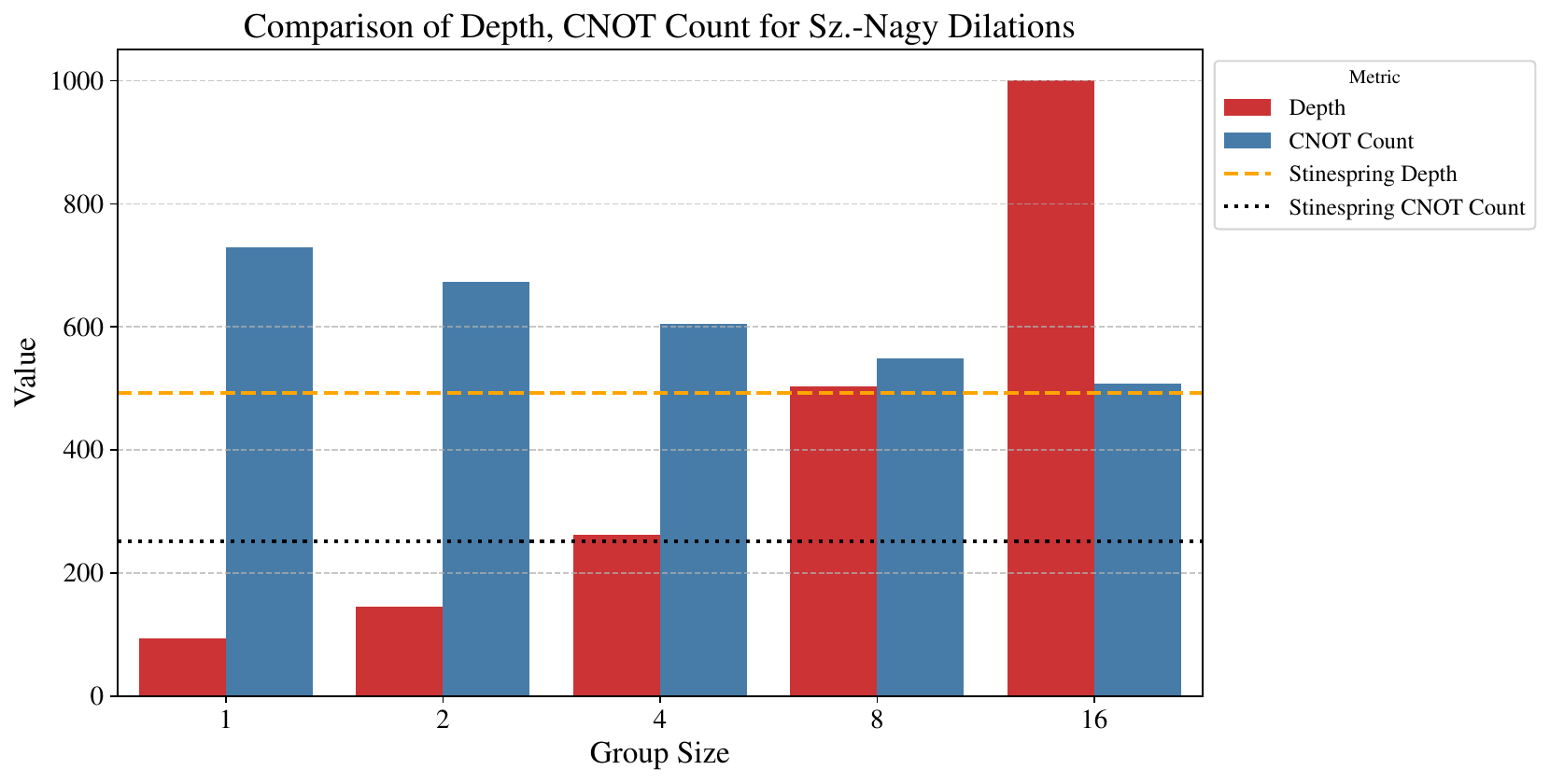}
    \caption{Comparison of Depth and CNOT count for the Sz.-Nagy dilations for different group sizes $\group$. The Depth for each group size is in red, while the CNOT count is in blue. The dashed/dotted orange/black line corresponds to the Stinespring dilation depth/CNOT count. }
    \label{fig:compare_groupsize}
\end{figure}

The comparison of the circuit depth and the CNOT count for different group sizes, using the Sz.-Nagy dilation, on a random set of $\nkraus = d^2$ Kraus operators acting on $2$ qubits can be seen in Fig.~\ref{fig:compare_groupsize}. The dotted/dashed lines correspond to the Stinespring dilation CNOT count/depth and are used as a reference. We observe that, as the group size $\group$ increases, the CNOT count decreases and the depth increases, demonstrating the trade-off between the two metrics. Additionally, we note that at $\group=\nkraus$ the Sz.-Nagy dilation has additional terms not present on the Stinespring dilation, so it is always better to utilize the Stinespring dilation in this case.

  \par We further explore the limit cases $l = 1, l = m$ and the effects on the dilation and mixing steps. When $l = m$ (which corresponds to a Stinespring-type dilation), we have a higher circuit depth, corresponding to the higher costs in dilation, but a low CNOT cost, as the mixing step cost is minimal. When $l$ decreases from $m$ to $1$, it increases the CNOT cost linearly and decreases the circuit depth logarithmically, until it reaches the Section~\ref{sec:combining_dilations} results, which can be interpreted as the lowest cost in implementing the dilations, and the highest cost in the mixing step. 
  \par The intermediate cases form a parametric family of dilations that gives us a choice of how much CNOT cost we need to gain to reduce the circuit depth by a certain amount. It is also interesting that these two block-encoding techniques (Stinespring and Sz.-Nagy) are two cases of the same dilation scheme.
  \par If we take the parameter CNOT/Depth, which is of order $O\left(\dfrac{ld^2}{d^2+\log(m) -\log(l)}\right)$, its minimum is attained at $l_{min} = me^{d^2+1} > m$, and for $l \in (0,l_{min})$ the function decreases. Therefore, for our method, the optimal technique to have the least number of CNOTS per unit of depth is when $l = m$, corresponding to the Sz.-Nagy combination.

%% file: experiments.tex
\section{Computational experiments}
\label{sec:computational_analysis}
In this section, we perform computational experiments to analyze the circuit depth, number of qubits, and CNOT counts of the simulation of open quantum systems using Stinespring dilation and the combined Sz.-Nagy and SVD dilations with groups of size $\group=1, 2$. 
The implementations were done in Python, using the Qiskit SDK. The code is available in a public repository~\cite{qclib}. 

Now, we compare the performance of each method when dealing with a random set of Kraus operators that act on a Hilbert space of $n$ qubits. We iterate on the number of qubits of the Hilbert space $n$ and randomly generate a set of $\nkraus = d^2 = 2^{2n}$ Kraus operators to be implemented by each method, as can be seen in Table~\ref{tab:methods_comparison_2}.
\begin{table}[H]
\centering
\begin{tabular}{|l|c|c|c|}
\hline
 \textbf{Methods} & \textbf{Circuit Depth} & \textbf{Number of CNOTs} & \textbf{Number of Qubits} \\ \hline
 Mixed SVD ($\group=1$)            & 69             & 521             & 93                       \\ \hline
 Mixed Sz.-Nagy ($\group=1$)            & 94             & 729             & 93                       \\ \hline
 Mixed SVD ($\group=2$)   & 122             & 593              & 60                       \\ \hline
 Mixed Sz.-Nagy ($\group=2$)   & 146	             & 673              & 60                       \\ \hline
 Stinespring      & 493             & 251             & 6                       \\ \hline

\end{tabular}
\caption{Comparison of different methods for implementing a full set of random Kraus operators that act on a Hilbert space of $2$ qubits. For the mixed (or combined) Sz.-Nagy and SVD algorithms $(\group = 1,2)$, we utilize one ancilla for each CSWAP implemented.}
\label{tab:methods_comparison_2}
\end{table}

\begin{table}[H]
\centering
\begin{tabular}{|l|c|c|c|}
\hline
 \textbf{Methods} & \textbf{Circuit Depth} & \textbf{Number of CNOTs} & \textbf{Number of Qubits} \\ \hline
Mixed SVD ($\group=1$)           & 168             & 5281             & 508                       \\ \hline
Mixed Sz.-Nagy ($\group=1$)           & 305             & 9761             & 508                       \\ \hline
Mixed SVD ($\group=2$) & 422             & 7417              & 315                       \\ \hline
Mixed Sz.-Nagy ($\group=2$) & 569             & 9561              & 315                       \\ \hline
Stinespring      & 8246             & 4145             & 9                       \\ \hline

\end{tabular}
\caption{Comparison of different methods for implementing a full set of random Kraus operators that act on a Hilbert space of $3$ qubits. For the mixed (or combined) Sz.-Nagy and SVD algorithms $(\group = 1,2)$, we utilize one ancilla for each CSWAP implemented.}
\label{tab:methods_comparison_3}
\end{table}

From Tables~\ref{tab:methods_comparison_2} and \ref{tab:methods_comparison_3}, we observe that the utilization of SVD dilations with $\group=1$ together with the CSWAPs for combining them yields the lowest circuit depth. However, it requires more circuit qubits than Stinespring dilation and SVD/Sz.-Nagy dilations with $\group=2$. SVD dilations also resulted in lower depth and CNOT count than Sz.-Nagy dilations, requiring the same number of qubits. The combined SVD/Sz.-Nagy dilations with $\group=2$ act as a balanced approach between the $\group=1$ and Stinespring methods and have a probability of $2/\nkraus$ compared to the probability of $1/\nkraus$ for $\group=1$. Therefore, choosing a grouping parameter $\group>1$ can be helpful depending on the available resources. For combining the outputs from the SVD/Sz.-Nagy dilations, we utilize one ancilla for each CSWAP implemented, leading to a lower depth but requiring a total of $(n+1)(\nkraus-1)$ ancillae. The results show how the circuit depth for Stinespring dilation grows extremely fast and quickly becomes unviable for NISQ devices~\cite{preskill2018quantum}.

%% file: fmosimnisq.tex
\section{Simulation of the Fenna-Matthews-Olson Dynamics on NISQ Devices}
\label{sec:FMO_real}
We will now show the feasibility and possible avenues for optimization in practical use cases for the previously introduced methods, focusing on the dynamics of the  Fenna-Matthews-Olson (FMO) Complex. 

In photosynthesis, the FMO complex acts in the transportation of excitons between the antennae and the reaction center\cite{engelEvidenceWavelikeEnergy2007a,scholesUsingCoherenceEnhance2017}. It has received intense investigation\cite{skochdopoleFunctionalSubsystemsQuantum2011,moixEfficientEnergyTransfer2011a,shabaniEfficientEstimationEnergy2012,panitchayangkoonDirectEvidenceQuantum2011,maiuriCoherentWavepacketsFenna2018,duanQuantumCoherentEnergy2022} due to the efficiency with which it transfers energy, which may be enabled by noise\cite{chinNoiseassistedEnergyTransfer2010}. By applying time discretization on the Lindblad master equation for the FMO Hamiltonian as in Ref.~\cite{Hu_2022_fmo}, we obtain the following Kraus operators
\begin{align*}
M_1 &= \sqrt{\alpha \delta t} |1\rangle\langle 1| & M_2 &= \sqrt{\alpha \delta t} |2\rangle\langle 2| \\
M_3 &= \sqrt{\alpha \delta t} |3\rangle\langle 3| & M_4 &= \sqrt{\beta \delta t} |0\rangle\langle 1| \\
M_5 &= \sqrt{\beta \delta t} |0\rangle\langle 2| & M_6 &= \sqrt{\beta \delta t} |0\rangle\langle 3| \\
M_7 &= \sqrt{\gamma \delta t} |4\rangle\langle 3| & M_0 &= \sqrt{\mathbf{I} - \sum_{k>0} M_k^{\dagger}M_k}.
\end{align*}
The constants used in the Kraus operators correspond to physical parameters of the FMO subsystem: $\alpha = 3\times 10^{-3}$ fs$^{-1}$ and $\gamma = 6.28\times 10^{-3}$ fs$^{-1}$ represent dephasing and relaxation rates, respectively; $\beta = 5\times 10^{-7}$ fs$^{-1}$ corresponds to the recombination rate to the ground state; and $\delta t = 48.4\, $ fs is the discrete timestep of simulation in femtoseconds (fs).

We analyze the costs of implementing the FMO dynamics in Table~\ref{tab:methods_comparison_fmo}. We compare the Stinespring dilation and the Sz.-Nagy/SVD dilations allied with the divide-and-conquer mixed-state preparation. The table lists the depth, CNOT count, and the number of qubits of each method.

\begin{table}[h!]
\centering
\begin{tabular}{|l|c|c|c|}
\hline
 \textbf{Methods} & \textbf{Circuit Depth} & \textbf{Number of CNOTs} & \textbf{Number of Qubits} \\ \hline
 Mixed SVD $(\group = 1)$          & 130             & 593             & 60                       \\ \hline
 Mixed  Sz.-Nagy $(\group = 1)$          & 250             & 847             & 60                       \\ \hline
 Mixed SVD $(\group = 2)$& 433             & 946              & 35                       \\ \hline
 Mixed Sz.-Nagy $(\group = 2)$& 489             & 942              & 35                       \\ \hline
 Stinespring      & 993             & 505             & 6                       \\ \hline

\end{tabular}
\caption{Comparison of different methods for implementing the described FMO operators. For the mixed (or combined) Sz.-Nagy and SVD algorithms $(\group = 1,2)$, we utilize one ancilla for each CSWAP implemented.}
\label{tab:methods_comparison_fmo}
\end{table}

To further discuss the feasibility of our proposed technique, we have simulated the FMO dynamics on the IBM-Kyiv\cite{javadi-abhariQuantumComputingQiskit2024} quantum device and compared its results with the expected statistical outcome. The chosen initial state was $\ket{\psi(0)} = \frac{1}{\sqrt{3}}[\ket{001}+\ket{010}+\ket{100}]$, which represents one possible state for energy transfer.

\begin{figure}[H]
    \includegraphics{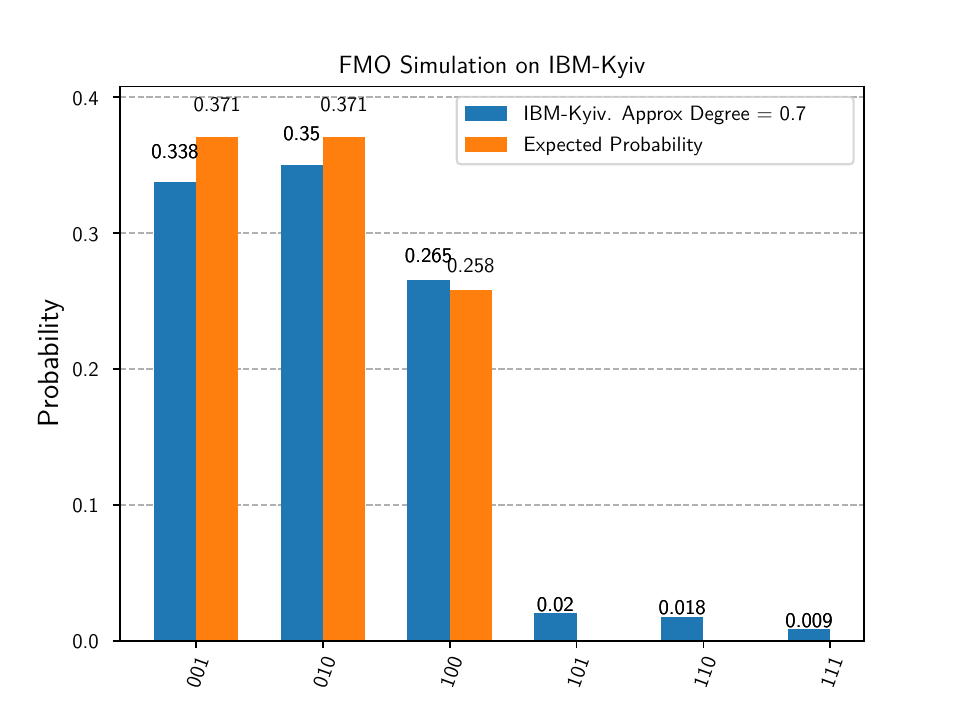}
    \caption{IBM Kyiv Implementation of FMO Operators with Approximate Degree = 0.7}
    \label{fig:kyiv_approx}
\end{figure}

However, due to the IBM-Kyiv quantum device topology, the routing and layout stage\cite{zouLightSABRELightweightEnhanced2024,liTacklingQubitMapping2019} added significant costs to the implementation as a whole. However, we retained fidelity in the actual results by employing heuristic approximations, as shown in Fig.~\ref{fig:kyiv_approx}.

\begin{figure}[H]
    \includegraphics{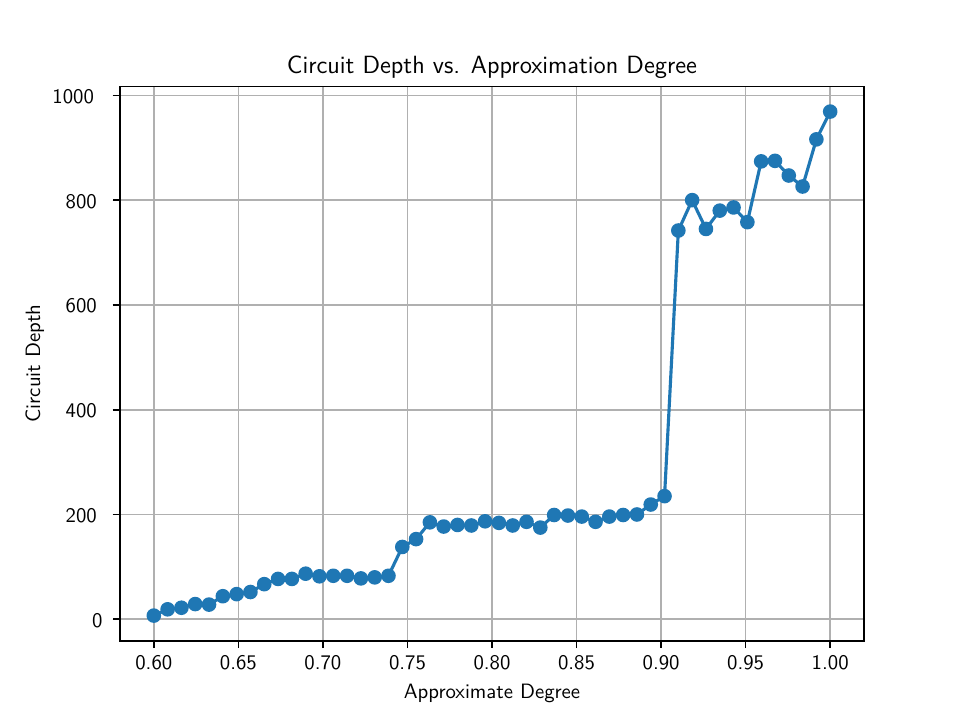}
    \caption{Comparing varying approximate degree values and Circuit Depth for the IBM Kyiv Backend.}
    \label{fig:approx_degreevsdepth}
\end{figure}

The Approximation Degree\cite{crossValidatingQuantumComputers2019}, which was the chosen heuristic approximation, acts as a multiplicative factor that affects the perceived fidelity of the back-end basis gates. As a consequence, it results in a trade-off between fidelity and circuit cost during synthesis, which can be used to minimize circuit depth, exemplified in Fig.~\ref{fig:approx_degreevsdepth}
In practical terms, the approximate degree merged and eliminated Kraus operators that did not substantially affect the dynamics for the proposed initial state. The optimization of the parameters above resulted in the Approximation Degree $0.7$ 

\par This case study shows that our proposed method can substantially optimize specific use cases, due to the capability of choosing which Kraus operators are relevant given a particular subset of initial states. Algorithms balancing out circuit fidelity, success rates, and circuit depth for specific maps remain to be explored in further work.

%% file: mt_cswap.tex
\section{Multi-target CSWAPs without ancilla qubits}
\label{sec:mtcswap}

\begin{figure}[ht]
    \centering
    \includegraphics[scale=1.2]{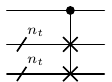}
    \caption{A Multi-target CSWAP, where there are two registers with $n_t$ qubits, and each pair of qubits is swapped. The control qubit controls the operation.}
    \label{fig:mt-cswap}
\end{figure}

\begin{figure}[htb]
    \centering
    \includegraphics[scale=1.2]{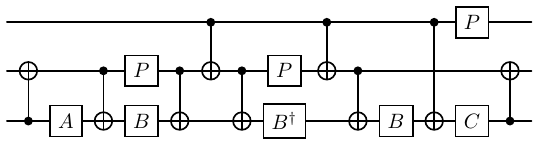}
    \caption{A controlled-swap gate, where the Toffoli gate is decomposed with the $C^2 U$, $U\in U(2)$ decomposition from Ref.~\cite{barenco_1995}. This decomposition only has $3$ CNOT gates involving the control qubit. For the $U=X$ gate, the values of the operators are $A=R_z (\pi/2)$, $B=R_y(-\pi/4)$, $C=R_z(\pi/2)R_y(\pi/4)$, and $P=T$ gate.}
    \label{fig:cswap-decomposition}
\end{figure}

In this section, we utilize the decomposition of a multi-target multi-controlled $X$ gate from Ref.~\cite{zindorf2024efficient} to construct a multi-target controlled-SWAP (CSWAP) without ancilla qubits. The method described here 
does not use ancilla qubits like Ref.~\cite{spacetimestateprep2023} and has a reduced constant in the leading term compared to other ancilla-free approaches~\cite{multitarget2018}. A CSWAP gate can be constructed with $2$ CNOTs and a Toffoli gate, and the multi-target CSWAP gate is the extension of the CSWAP to the case of multi-qubit target registers, shown in Fig.~\ref{fig:mt-cswap}. As the only CNOT gates that cannot be parallelized are the ones involving the control qubit of the CSWAP, we have chosen a Toffoli decomposition that minimizes this parameter. The resulting CSWAP decomposition can be seen in Fig.~\ref{fig:cswap-decomposition}, where we utilize the decomposition from Ref.~\cite[Lemma 6.1]{barenco_1995}, but only apply some of the optimizations present in Ref.~\cite[Corollary 6.2]{barenco_1995}~\cite[Lemma 5]{iten2016quantum}, resulting in only $3$ CNOT gates that involve the control qubit. The gate is decomposed into CNOTs and unitary gates and has a CNOT count of $9$ and a total depth of $14$.

So, for a multi-target CSWAP gate, the only operations that cannot be executed in parallel are the $3$ CNOT gates which utilize the control qubit of the CSWAP. Each $3$ controlled $X$ gates will act on one of the qubits of each pair of targets, corresponding to  $3$ $\ntargets$-targets CNOT gates. Using the result of Ref.~\cite{zindorf2024efficient}, a controlled $X$ gate for multiple targets can be implemented with a depth of $1+2 \lceil \log_2 \ntargets \rceil$. This sequence of arguments leads us to the following theorem:

\begin{theorem}
A controlled-swap gate with $\ntargets$ pairs of targets can be implemented with a depth of $6 \lceil \log_2 \ntargets \rceil+14$.
\end{theorem}